\documentclass[9pt,twocolumn,twoside]{osajnl}

\journal{ol} 

\setboolean{shortarticle}{true} 

\title{Two-color interpolation of absorption response for quantitative Acousto-Optic imaging}

\author[1,*]{Ma\"imouna Bocoum}
\author[1]{Jean Luc Gennisson}
\author[1]{Caroline Venet}
\author[2]{Mingjun Chi}
\author[2]{Paul Michael Petersen}
\author[3]{Alexander A. Grabar}
\author[1]{François Ramaz}

\affil[1]{Institut Langevin, Ondes et Images – ESPCI Paris, PSL Research University, CNRS UMR 7587, INSERM U979, Université Paris VI Pierre et Marie Curie, 1 rue Jussieu, 75005 Paris, France}
\affil[2]{DTU Fotonik, Department of Photonics Engineering, Technical University of Denmark, Frederiksborgvej 399, P.O. Box 49, DK-4000 Roskilde, Denmark}
\affil[3]{Uzhgorod National University, Institute of Solid State Physics and Chemistry, 88000, Uzhgorod, Voloshyn st. 54, Ukraine}

\affil[*]{Corresponding author: physics@mbocoum.fr}

\ociscodes{(110.4234) Multispectral and hyperspectral imaging; (110.5120) Photoacoustic imaging;(110.1085) Adaptive imaging; (110.7170) Ultrasound; (290.4210) Multiple scattering.}

\begin{abstract}
Diffuse Optical Tomography (DOT) is a reliable and widespread technique for monitoring qualitative changes in absorption inside highly scattering media. It has been shown, however, that Acousto-Optic (AO) imaging can provide significantly more qualitative information, without the need for inversion algorithms, due to the spatial resolution afforded by ultrasound probing. In this article, we show how, by using multiple-wavelength AO imaging, it is also possible to perform quantitative measurements of absorber concentration inside scattering media.
\end{abstract}

\setboolean{displaycopyright}{true}

\begin{document}

\maketitle

\section{Introduction}

 The development of non-invasive imaging techniques for accessing the absorption and scattering properties of biological tissues in the near-infrared (NIR) range is critical to multiple medical applications, such as early cancer diagnosis. For example, in the $700$-$800\,\mathrm{nm}$ spectral window, the reduced scattering coefficient $\mu_s'$ of healthy breast tissue is $\sim 8~\mathrm{cm}^{-1}$, and the absorption coefficient is $\mu_a\sim 0.04~\mathrm{cm}^{-1}$~\cite{durduran2002bulk}. Tumors induce local variations in blood concentration and oxygen level, such that a malignant \textit{ductal carcinoma} will result in $\mu_a$ being two to four times higher and $\mu_s'$ nearly 20\% higher than that of healthy breast tissue, with typical sizes ranging from hundreds of microns to several centimeters~\cite{grosenick2005time,fantini1998assessment,holboke2000three}, depending on the state of the tumor. Moreover, a local variation in the optical absorption of tissues over time can be exploited to monitor any changes in chemical composition, as in Hb/HbO oxymetry measurements~\cite{boas2001imaging}. Indeed, the Beer-Lambert law, $\mu_a(\lambda) = \sum_{i}\ln(10)\epsilon_i(\lambda) [C]_i$ with $\epsilon_i$ the extinction coefficient per distance unit, provides a straightforward relation between the absorption coefficient $\mu_a$ at a given point and the local absorber concentrations  $[C]_i$. Measuring $\mu_a$ at different wavelengths enables one to invert the above relation and recover the unknown concentrations. When performing a global measurement involving multiple light scattering, the main difficulty lies in estimating $\mu_a$ locally. This problem is the main focus behind NIR Spectroscopy~\cite{boas2001imaging}, and one frequently encountered problem lies in establishing the relation between $\mu_a$ in a given region of interest (ROI) and the measured signal. Diffuse Optical Tomography (DOT)~\cite{boas2001imaging} based on single or multiple spatial, temporal or spectral detection relies on inversion algorithms to provide information on the optical properties of scattering tissues~\cite{chance1988comparison,sevick1991quantitation,tromberg2000non,choe2005diffuse}. Because of a limited level of accuracy, however, they fail to provide trustworthy images of tissue absorption, especially from ROI at high penetration depths~\cite{gunadi2011spatial}. Some groups have managed to retrieve quantitative information either by measuring apparent optical paths inside the scattering media and thereby using a corrected Beer-Lambert Relation~\cite{wray1988characterization} or relying on an increased number of illumination sources and detectors to run inversion algorithms~\cite{patterson1989time}. 
\noindent Acousto-Optical Tomography (AOT)~\cite{marks1993comprehensive,wang1995continuous,hawrysz2000developments,resink2012state} was introduced twenty years ago as a way to improve the optical contrast inside scattering media. It is complementary to pure ultrasound imaging (US) that only reveals pronounced structural discontinuities inside tissues. In AOT, a controlled MHz-acoustic wave interacts with coherent light, inducing a local shift in the light carrier frequency through the acousto-optic (AO) effect~\cite{wang2001mechanisms}. Because ultrasounds (US) propagate ballistically inside soft tissues, the so-called  "tagged" photons inherit the spatio-temporal resolution of the controlled US. The tagged photons are then detected using either single-detection autocorrelation~\cite{lev2000ultrasound}, parallel speckle contrast imaging~\cite{leveque1999ultrasonic,li2002methods}, spectral hole burning~\cite{li2008pulsed} or, as exposed in the present work, self-adaptive wavefront holography using a photorefractive crystal~\cite{ramaz2004photorefractive,murray2004detection}.
\noindent AO imaging could be considered as an extension of NIRS imaging, where an extensive number of virtual sources are placed in a given plane, which drastically increases our knowledge of the system. It is often compared to photoacoustic (PA) tomography~\cite{wang2004ultrasound}, another imaging technique that relies on the detection of ultrasonic waves generated by local optical absorption of tissues~\cite{van2015review}. In both modalities however, the unknown bulk optical properties of the scattering media hinders quantitative imaging. Little work has been done on quantitative AO imaging and to our knowledge, only one group has performed a two-color quantitative measurement of a dye mixture using pure AO measurement~\cite{kim2007multi}. This measurement however, was performed assuming a Beer-Lambert-like relation between the AO signal and the actual absorption coefficient $\mu_a$. On the other hand, several solutions have been proposed in PA imaging, which do not rely on the propagation equation, but rather on the signal decomposition into a limited number of components, thereby turning to sparse representation of the measured data~\cite{chen2001atomic}. For example, an elegant quantitative absorption measurements was done by discriminating the fast and slow spatial variations of absorption using principal component decomposition~\cite{rosenthal2009quantitative}, or a two-color quantitative measurement shown to be when the bulk properties of the medium are known~\cite{zemp2010quantitative}. Another group proposed to couple PA and AO imaging to eliminate the integrated influence of the propagation medium from the ROI to the detector~\cite{daoudi2012correcting,hussain2016quantitative}.

\noindent Which ever the approach, quantitative retrieval is possible when the system's internal degrees of freedom does not exceed the number of measured data points. Here, we show how the use of multiple wavelengths allows, for a given geometry, to gain sufficient information about the dependence of the AO signal when varying the absorption coefficient $\mu_a$. As a consequence, we demonstrate the possibility to quantitatively measure the relative concentration variations of a chromatic absorber inside a scattering medium when the Beer-Lambert law is not applicable.\\

\begin{figure}[htbp]
\centering
\includegraphics[width=\linewidth]{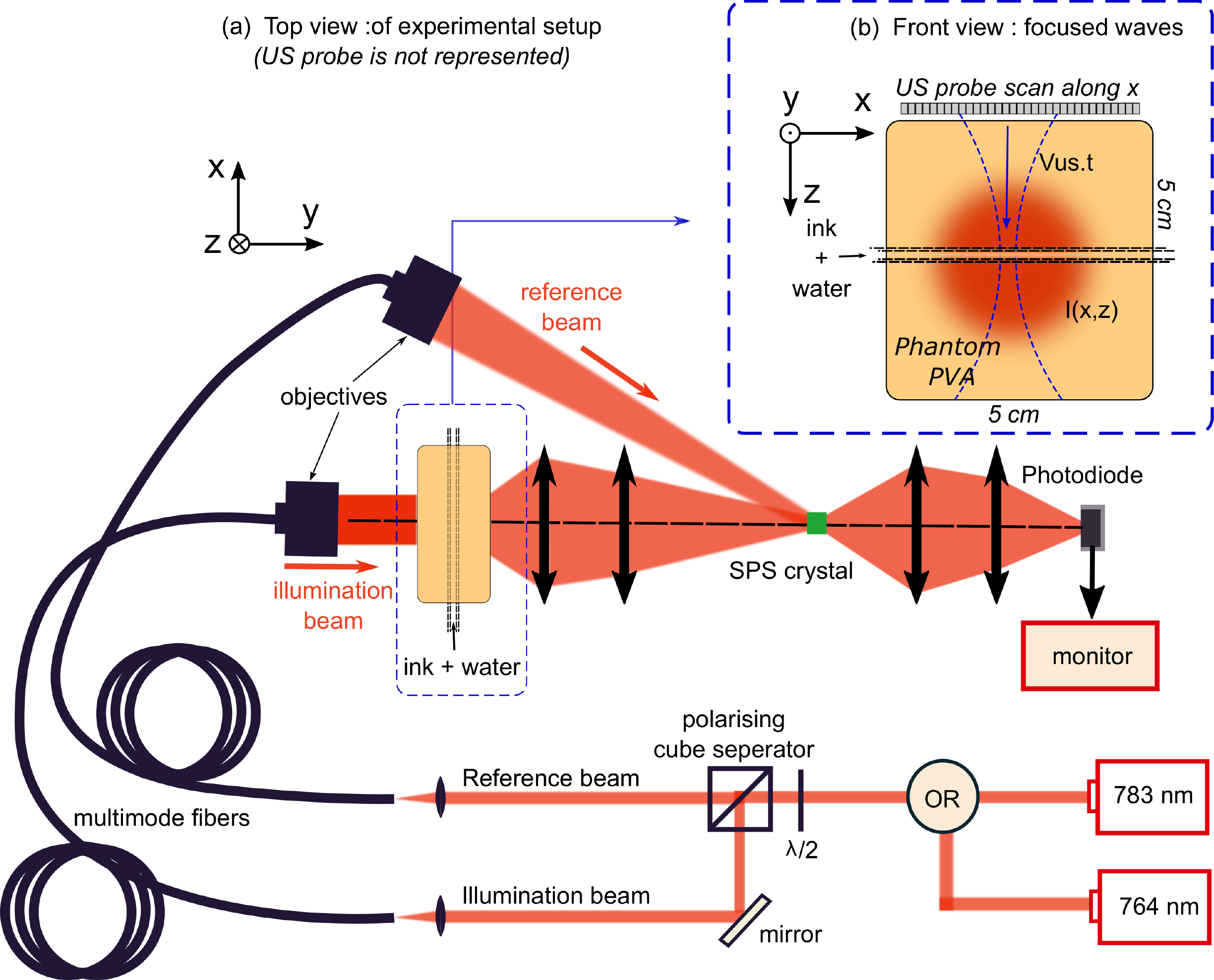}
\caption{(a) Experimental setup. OR: flip mirror to choose between $764\,\mathrm{nm}$ or $783\,\mathrm{nm}$. The beam is separated with a polarizing cube and each replica coupled into a $200\,\mu\mathrm{m}$ inner diameter multi-mode fiber. One collimated fiber output illuminates the phantom, and the transmitted scattered light is collected by a set of lenses onto the photorefractive crystal (PRC). From the other fiber, a reference beam is focused onto the PRC for wavefront holography at $\sim 50\,\mathrm{mW/cm^2}$  (b) Schematic front view of the PVA phantom. A plastic tube of $2.5\,\mathrm{mm}$ inner diameter is inserted and filled with China ink of variable concentration. The US probe will tag photons in the $(xz)$ plane containing the tube.}
\label{fig:setup}
\end{figure}

AO imaging is sensitive both to scattering and absorption. To study the influence of local variations only in absorption, a HxLxW = $5\mathrm{x}5\mathrm{x}4\mathrm{cm}$ phantom was made out of 10\% Polyvinyl alcohol phantom (PVA) with $\mu_s'\sim 6\mathrm{cm}^{-1}$~\cite{boutet2011fantome}. A plastic tube was inserted in the transverse direction $x$, as shown in Fig~\ref{fig:setup}(b), and filled with China ink diluted in a variable volume of water such as to vary its absorption coefficient $\mu_a$. The full experimental setup is shown in Fig~\ref{fig:setup}(a). Using a flip mirror, we illuminated the sample either with a single mode Master Oscillator Power Amplifier (MOPA, \textit{Sacher Lasertechnik GmBH}) System at $\lambda = 764\,\mathrm{nm}$ or with an external-cavity diode laser system developed by \textit{Norlase} and \textit{DTU Fotonik} at $\lambda= 783\,\mathrm{nm}$~\cite{chi2005tunable}. A commercial ultrasonic probe (SL10-2,0.2mm pitch from \textit{Supersonic Imaging}) was used to image the tube in the plane orthogonal to the optical axis. Photons scattering through the phantom were collected by a set of two lenses and refocused on a Te-dopped $\mathrm{Sn}_2\mathrm{P}_2\mathrm{S}_6$ photorefractive crystal  (SPS:Te1\%), in negative gain configuration~\cite{farahi2010photorefractive,laudereau2015multi} where it undergone a two-wave mixing process, with a plane wave reference beam of $\sim 50\,\mathrm{mW/cm^2}$. The back surface of the crystal was then imaged onto a $S = 13\,\mathrm{mm^2}$ photodiode (\textit{Thorlabs PDA36 A}). The resulting voltage was filtered between $200\mathrm{Hz}$ and $2\,\mathrm{MHz}$ to keep only the AC component of the signal. \\

\begin{figure}[htbp]
\centering
\includegraphics[width=\linewidth]{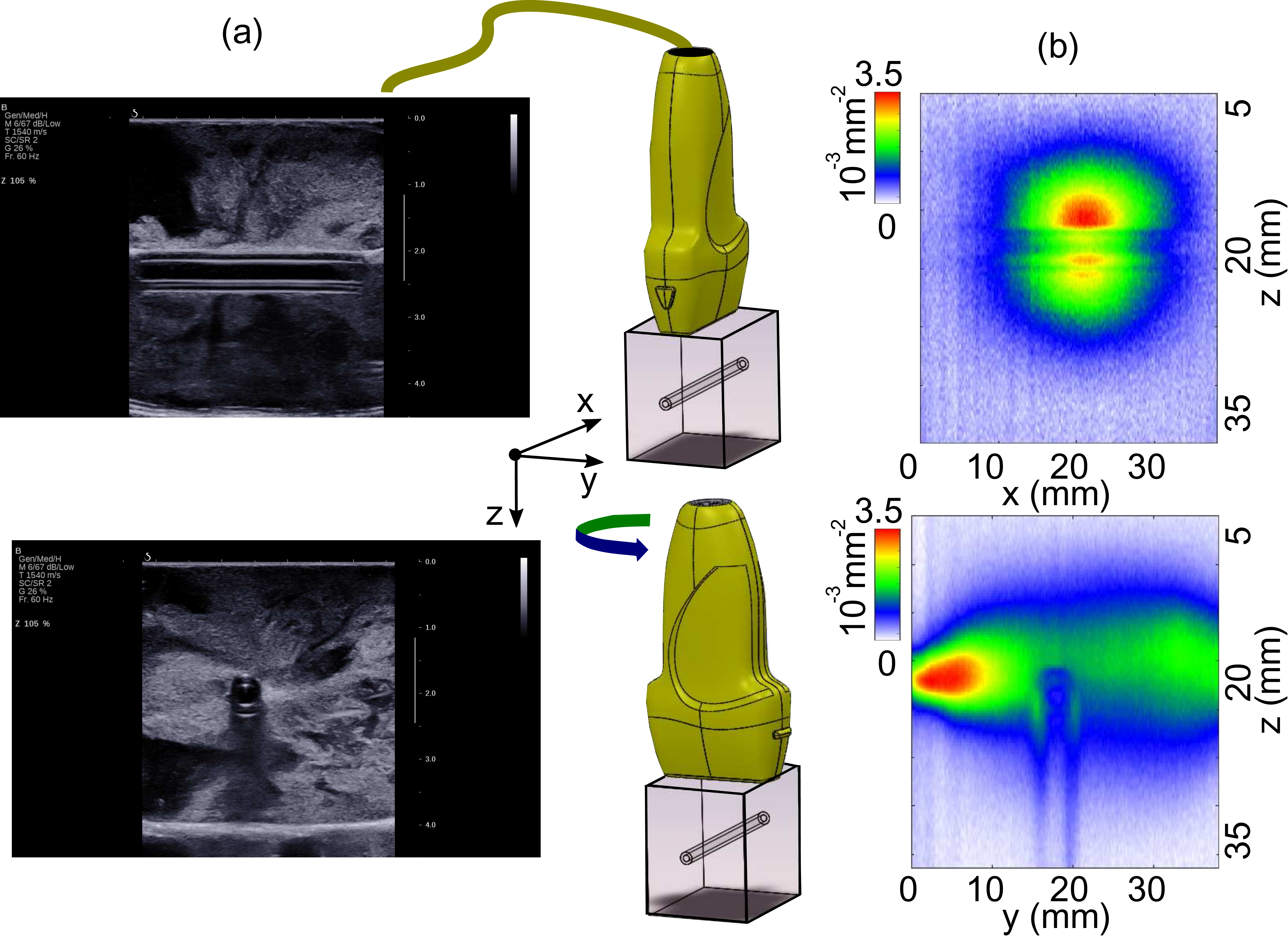}
\caption{(a) B-mode images of the PVA phantom obtained in clinical conditions where the acoustic probe was positioned respectively orthogonal to (top) and along (bottom) the optical axis (b) AO images in the same configurations.}
\label{fig:echograph}
\end{figure}

\noindent An image was then acquired as follow: for a given $x$ position, the acoustic probe generated a two-cycle focused wave centered at $3\,\mathrm{MHz}$ propagating along $z$. A synchronized 14-bit acquisition card (\textit{Gage Digitizer}) recorded the AO filtered signal. The $z$ coordinate was obtained by multiplying the acquisition time by the sound propagation in water $V_{us}\sim 1450\mathrm{m/s}$. 
This time-resolved focusing scan along dimension $x$ was repeated $1000$ times to improve signal to noise ratio through averaging. In Fig~\ref{fig:echograph}(b), we show the qualitative comparison between the AO image obtained when the probe is respectively orthogonal (top) and along the optical axis (bottom) . In both cases, the AO image is compared with that provided by a clinical ultrasound imaging device (Aixplorer,\textit{Supersonic Imaging}) in the same respective configurations. The resulting B-mode images shown in Fig~\ref{fig:echograph}(a) provide only structural information on the phantom. We used them to precisely locate the tube edges in the AO image and therefore define the  ROI corresponding to its inner volume. To conduct the experiment presented here, we picked the configuration where the probe is orthogonal to the optical axis (y), thereby choosing a larger tube's ROI to minimize the signal to noise ratio after integration. The tube ROI was defined by $18.9\,\mathrm{mm} \le z \le 22.5\,\mathrm{mm}$.

\noindent The normalized  ink concentration $[C]$ was varied from $0$ (pure water) to 1 (pure ink) and for each value, an AO image was taken at $\lambda_1 = 764\mathrm{nm}$ and $\lambda_2 = 783\mathrm{nm}$ respectively. To derive $\mu_a$ from the ink concentration $[C]$, we used the local relation:
\begin{equation}
\label{eq:dilrelation}
\mu_a(\lambda)  =\mu_{a,w}(\lambda) + [C]\mu_{s,0}(\lambda)  
\end{equation}
\noindent where $\mu_{a,w}$ and $\mu_{s,0}$  are respectively the absorption coefficient of water and that of the mother ink solution measured with an absorption spectrometer prior to the experiment. 

\begin{figure}[htbp]
\centering
\includegraphics[width=0.9\linewidth]{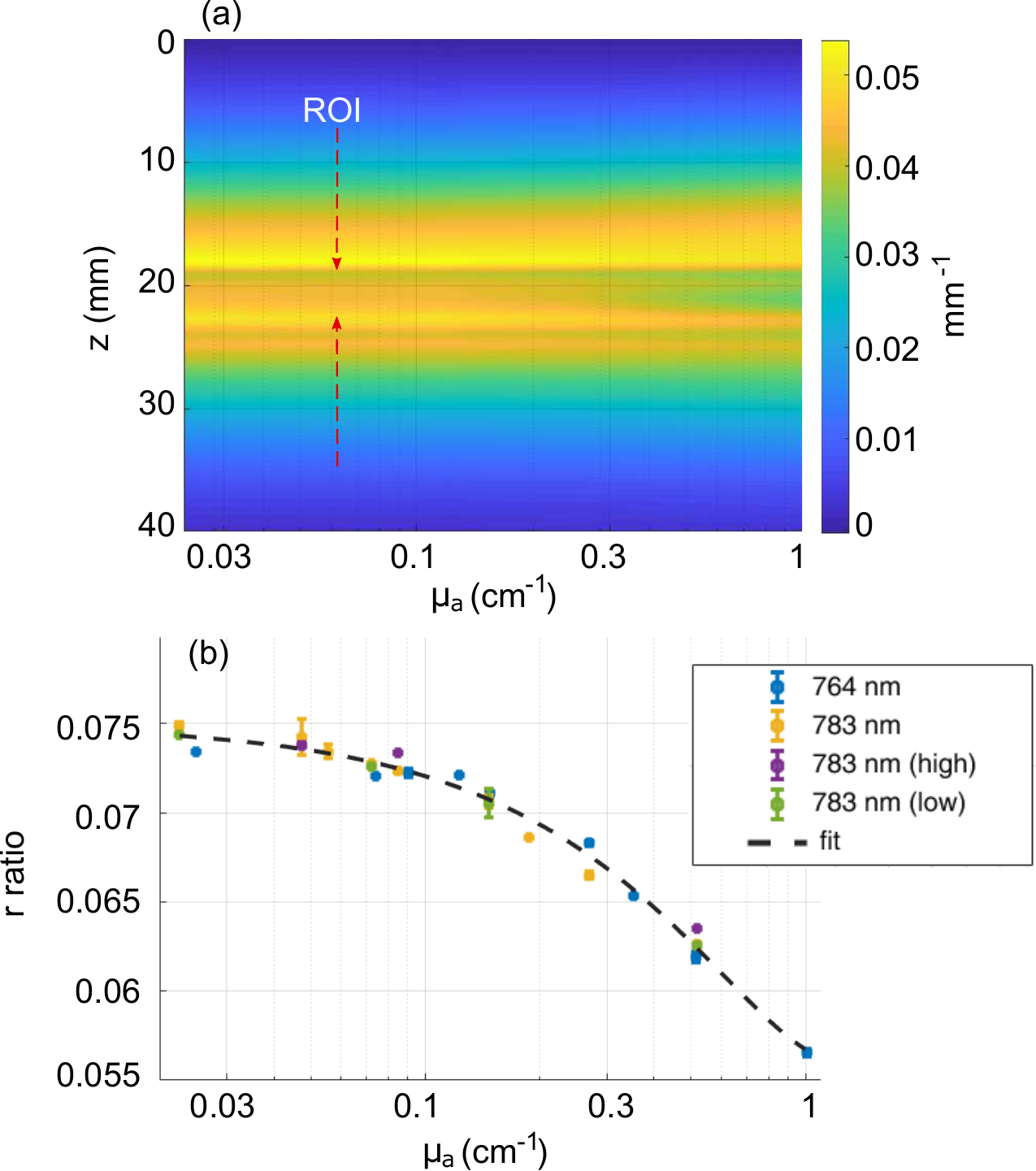}
\caption{ (a) Integrated signal over direction $x$ when changing the ink concentration, and thereby $\mu_a$, in the tube which ROI is delimited by the arrows (b) Measurement of ratio r defined by Eq~\ref{eq:ROI} with respect to $\mu_a$ at two wavelength and for higher and lower illumination power. The dotted black line is a quadratic fit on all this experimental data points.}
\label{fig:mua}
\end{figure}

 For each AO image I(x,z), we subtracted the mean background calculated from a region where the signal is obviously equal to zero. The image is then renormalized so that its integral over the whole profile is equal to one. The qualitative change in local absorption can be seen in Fig~\ref{fig:mua}(a), where the image integrated over direction $x$ is plotted with respect to $\mu_a$, for $\lambda = 764\mathrm{nm}$. Inside tube ROI, indicated by arrows, the drop in signal density is consistent with a stronger absorption as the ink concentration is increased. 
The ratio r between the signal fully integrated over the tube ROI and that contained in the rest of the image is defined by Eq~\ref{eq:ROI}: 

\begin{equation}\label{eq:ROI}
r =  \frac{\iint\limits_{(x,z)\in \mathrm{ROI}}I(x,z)dxdz }{\iint\limits_{(x,z)\notin \mathrm{ROI}}I(x,z)dxdz}
\end{equation}

In Fig~\ref{fig:mua} (b), $r$ was measured at both $764\mathrm{nm}$ and $783\mathrm{nm}$ for an input power of $480\mathrm{mW}$, and plotted with respect to the absorption coefficient $\mu_a$. The result shows that $r$($\mu_a$) is independent of the illumination wavelength. To test the consistency of this result, we also measured $r$ at $783\mathrm{nm}$ for an input power of $650\mathrm{mW}$ (high) and $360\mathrm{mW}$ (low) respectively. The result shows the ratio r($\mu_a$) does not depend on illumination power. A fit (dotted line) performed on all data points shows that, in our geometrical configuration, $r$ depends quadratically on $\mu_a$.

\noindent We now propose to retrieve the relative variation in ink concentrations inside the tube with respect to $\mu_a$. For a given data range, $r$ can be decomposed in a series:

\begin{equation}
\label{eq:series} 
r(\mu_a) = \sum_{0}^{N}\alpha_n \mu_a^n
\end{equation}

\noindent In general, $N$ will be equal to $\infty$. However, for a given $\mu_a$ interval a finite number of terms is usually sufficient to properly approximate the function. Within the scope of our experiment where $0.02\le\mu_a\le 1$, we saw that $N=2$, which allows us to write:

\begin{equation}
\label{eq:S}
r(\mu_a) = \alpha_2 \mu_a^2 + \alpha_1  \mu_a + \alpha_0 
\end{equation}

\noindent $\alpha_0$ is the signal level when $[C] = 0$, so it corresponds to the signal level when only water is inserted in the tube. We can therefore redefine the measured signal as  $ r(\mu_a) \leftarrow r(\mu_a)  - r(\mu_{a,w}) $ and set $\alpha_0$ to zero. The remaining polynomial coefficients $\alpha_1$ and $\alpha_2$ are the two degrees of freedom of $r$. We propose to retrieve them using a two-color measurement at a known concentration $[C]_0$, therefore matching the number of measured data points with the system's internal degrees of freedom. Injecting Eq~\ref{eq:dilrelation} into Eq~\ref{eq:S}, the signals $r(\lambda_1)$ and $r(\lambda_2)$ are related to the polynomial coefficients as follows:

\begin{equation}
\label{eq:matrice}
\begin{pmatrix}
r(\lambda_1)\\
r(\lambda_2)
\end{pmatrix}
=
\begin{pmatrix}
(\mu_{1}[C]_0 + \mu_{1,w}) & (\mu_{1}([C]_0 +\mu_{1,w} )^2 \\
(\mu_{2}[C]_0 + \mu_{2,w}) &  (\mu_{2}[C]_0 +\mu_{2,w} )^2
\end{pmatrix}
\begin{pmatrix}
\alpha_1\\
\alpha_2 
\end{pmatrix}
\end{equation}

\noindent Where $\mu_{1} = \mu_{a,0}(\lambda_1)$ and  $\mu_{2}= \mu_{a,0}(\lambda_2)$ are the known absorption coefficients of the ink at respectively $\lambda_1$ and $\lambda_2$, while $\mu_{1,w} =\mu_{a,w}(\lambda_1) $ and $\mu_{2,w} =\mu_{a,w}(\lambda_2) $ are that of pure water. We retrieved $\alpha_1$ and $\alpha_2$ by inverting Eq~\ref{eq:matrice}, which means $r(\mu_a)$ is now fully defined. 

\begin{figure}[htbp]
\centering
\includegraphics[width=0.9\linewidth]{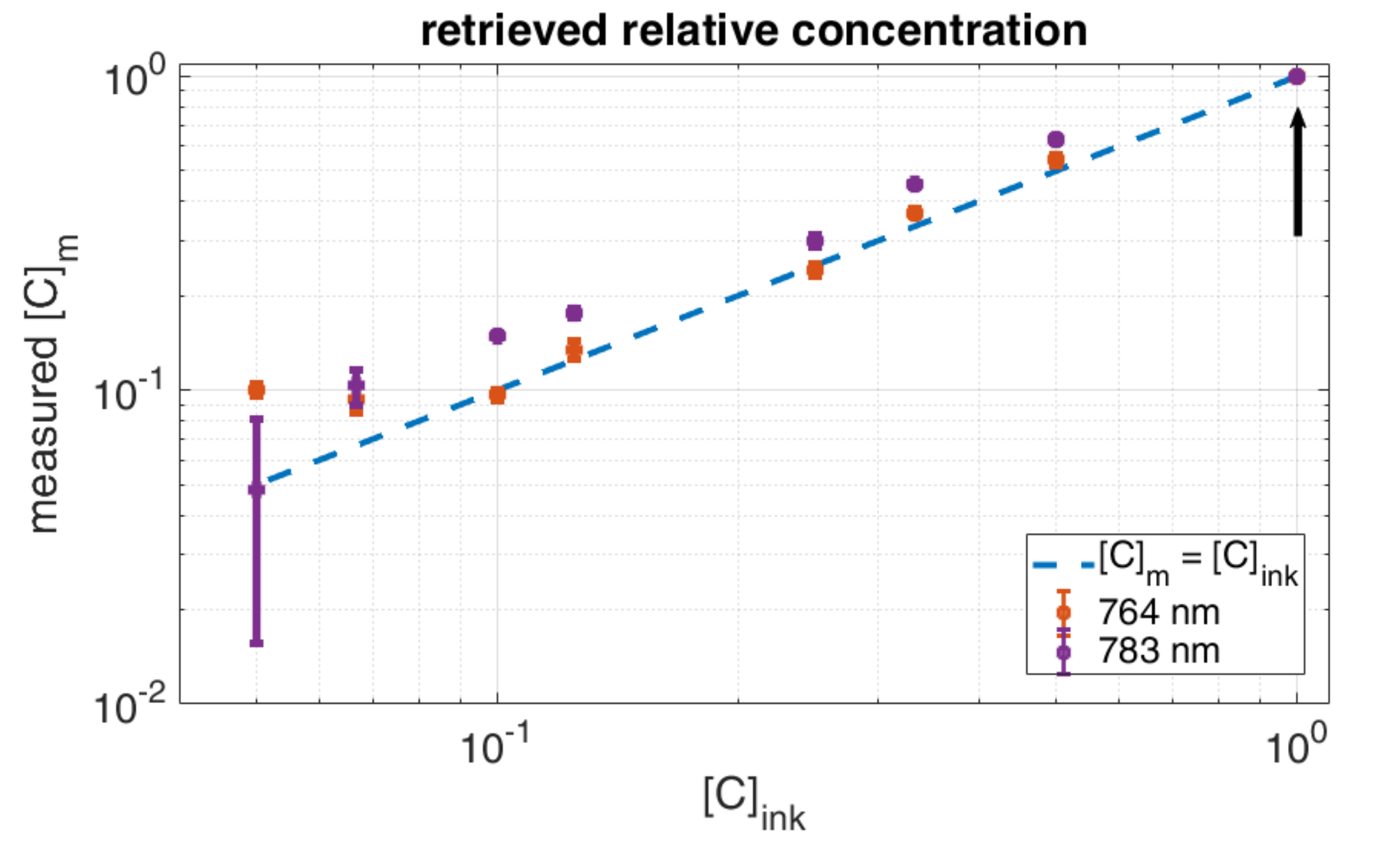}
\caption{ Measured ink concentration $[C]_\mathrm{m}$ with respect to actually known concentration $[C]_{\mathrm{ink}}$ for data points at $764\,\mathrm{nm}$ and $783\,\mathrm{nm}$ after two-color interpolation. The arrow indicates the reference point $[C]_0$ from which the polynomial coefficients $\alpha_1$ and $\alpha_2$ were calculated.}
\label{fig:retrieval}
\end{figure}

\noindent The inversion could in principle be performed at any concentration $[C]_0$, so we choose the highest concentration $[C]_0 = 1$ as our reference point, with the strongest chromatic sensitivity in absorption. Using the reconstructed response $r$, it is quite straightforward to recover the ink concentration from the AO signal at either $764\mathrm{nm}$ or $783\mathrm{nm}$. The results are presented in Fig~\ref{fig:retrieval} where we observe a good match between the measured values $[C]_{\mathrm{m}}$ and actual input values $[C]_{\mathrm{ink}}$ indicated by the dotted line, with a $25\%$ error on average. Note that the reconstruction is less accurate when $[C]\sim 0$, where we reach a limit in differentiation between two ink concentrations.

In conclusion, we have demonstrated the possibility of  measuring the relative concentration of ink inside a scattering medium through a two-color interpolation of the AO response within a given region of interest.  In principle, one could always develop the system response  into a polynomial sum with an infinite number of terms. If only $n$ terms are necessary to approach the response, then $n$ wavelengths are needed to calculate the system response and therefore perform quantitative measurements relative to a known value (absorption coefficient of healthy tissues, for instance). Of course, implicit hypotheses are made in order for this reconstruction to be valid, namely (i) the reduced scattering coefficient $\mu_s'$ is independent of ink concentration and (ii) the bulk scattering properties vary very little with the illumination wavelength. The latter is necessary to eliminate the spectral response of the background and perform a proper calibration of the AO response before following the evolution of chromophore concentrations. When this approximation is no longer valid, accounting for the background response can be conceived as an extension of the internal degrees of freedom of the AO response, and compensated by performing the measurement at an increased number of wavelengths. This is essentially the idea behind specific PA imaging of chromophores using spectral unmixing~\cite{glatz2011blind}.  This article therefore emphasizes the need for tunable NIR laser sources compatible with AO detection requirements, such as high mode stability. Such multiple-wavelengths AO systems would enable quantitative imaging inside scattering media.

\section{Funding Information}

With financial support from ITMO Cancer AVIESAN (Alliance Nationale pour les Sciences de la Vie et de la Santé, National Alliance for Life Sciences \& Health) within the framework of the Cancer Plan under contract C16027HS

\bibliography{ReferencesList}

\end{document}